\documentclass[prb,twocolumn,superscriptaddress,amsmath,amssymb,showpacs]{revtex4}
\usepackage{graphicx}
\usepackage{bm}

\begin{document}
\title{Dynamic nuclear spin polarization in self-assembled quantum dots \\
under zero magnetic field}

\author{R.\ Matsusaki}
	\affiliation{Division of Applied Physics, Hokkaido University, N13 W8, Kitaku, Sapporo 060-8628, Japan}
\author{R.\ Kaji}
\email{r-kaji@eng.hokudai.ac.jp}
	\affiliation{Division of Applied Physics, Hokkaido University, N13 W8, Kitaku, Sapporo 060-8628, Japan}	
\author{S.\ Yamamoto}
	\affiliation{Division of Applied Physics, Hokkaido University, N13 W8, Kitaku, Sapporo 060-8628, Japan}
\author{H.\ Sasakura}
	\affiliation{Division of Applied Physics, Hokkaido University, N13 W8, Kitaku, Sapporo 060-8628, Japan}	
\author{S.\ Adachi}
\email{adachi-s@eng.hokudai.ac.jp}
	\affiliation{Division of Applied Physics, Hokkaido University, N13 W8, Kitaku, Sapporo 060-8628, Japan}

\date{\today}

\begin{abstract}
We studied the zero-field dynamic nuclear spin polarization in a single In$_{0.75}$Al$_{0.25}$As/Al$_{0.3}$Ga$_{0.7}$As quantum dot. Even without any external magnetic field, the positive trion excited by the circularly-polarized light generated an Overhauser field of up to $\sim$0.8 T. From the excitation power dependences of the Overhauser field and degree of circular polarization of the photoluminescence spectra, the relation between the Overhauser field and Knight field under zero external magnetic field was revealed clearly. In addition, we found that the nuclear depolarization rate decreased as the magnitude of the longitudinal magnetic field increased, which seemed to be caused by the influence of the quadrupolar interaction of nuclear spins. Further, the key parameters describing the dynamics of a coupled electron-nuclear spin system, the electron g-factor and the fluctuation of the Overhauser field, were evaluated in a typical single InAlAs quantum dot.
\end{abstract}
\maketitle

\section{Introduction}\label{Intro}
The carrier spin dynamics in low-dimensional semiconductor structures has attracted considerable interest because of the possibilities of spin storage and manipulation in future semiconductor electro-optic devices and quantum information processing~\cite{ALS02,SpinPhysics,Slavcheva10,Taylor03,Muto05}. This is because the confined carrier spins are relatively well decoupled from the orbital and charge degrees of freedom and the spin coherence is not obstructed by the commonly noted charge decoherence mechanisms~\cite{Khaetskii00}.
For these potential applications, 
a complete understanding of the fundamental interactions among the spins in the localization volume is crucial as these interactions may limit the application performance because of the resultant spin decoherence. Hyperfine interaction (HFI), which is the magnetic interaction between the confined carrier spins and lattice nuclei, becomes especially important.

Because the lattice nuclei act as a reservoir for an optically or electrically injected electron spin, the \textit{engineering of nuclear spins} such as the optical manipulation of the nuclear spin polarization (NSP) not only leads to the potential applications but also opens up a new spin physics. Along this line, the dynamic nuclear spin polarization (DNSP) induced by the electron-nucleus flip-flop part (e-n FF) of HFI and the resulting Overhauser field have been investigated intensively in semiconductor quantum dots (QDs)~\cite{Gammon01,Yokoi05,Bracker05,Eble06,Urbaszek13}. Because of the strong localization of the carrier wave function in a QD, the role of HFI in spin dynamics is drastically enhanced compared with those in bulks and quantum wells. One of the direct consequences of this enhanced HFI is a severe relaxation of electron spin polarization caused by a fluctuation of the Overhauser field in QDs~\cite{SpinPhysics,Merkulov02,Braun05}. Similarly, the fluctuation of the Knight field, which is the electron-generated effective magnetic field seen by each nucleus, induces the nuclear spin relaxation~\cite{OptOrientation}. Interestingly, both effective fields also help to stabilize the spin counterpart under some conditions. Such a reciprocal influence makes the spin physics in nanostructures complicated and interesting. Until now, many fascinating and remarkable works have been reported: the bistability of the nuclear spin system~\cite{Braun06,Tartakovskii07,Maletinsky07,Kaji08}, locking of the QD transition energy to the excitation laser~\cite{Latta09,Xu09,Hogele12}, optically detected nuclear magnetic resonance (NMR) and spin echo for a single QD~\cite{Chekhovich12,Chekhovich15,Stockill16}, and so on.

In contrast to those studies performed under large external magnetic fields, experimental reports under zero magnetic field have so far been insufficient~\cite{Lai06,Oulton07,Belhadj09,Larsson11,CFFong16}.  
As shown later, excitation by a circularly polarized beam induces a considerable Overhauser field under a wide range of experimental conditions, even if one does not intend to induce an Overhauser field. Thus, the understanding of HFI is necessary to analyze the experimental results performed under zero external magnetic field. 
In addition, zero-field DNSP reveals valuable information about the nuclear quadrupolar effect, which has been recently of interest from the viewpoint of stabilizing the NSP under zero field~\cite{Dzhioev07,Krebs10} and revealing the novel spin dynamics related to noncollinear HFI~\cite{Hogele12,Huang10}.

In this work, we study zero-field DNSP in a single InAlAs QD by focusing on the photoluminescence (PL) spectra of a positive trion ($X^{+}$). The degree of circular polarization (DCP) of $X^{+}$-PL works as a direct measure of the average electron spin polarization $\langle S_{\rm z} \rangle$. Even without any external magnetic field, the spin-selected electron with circularly polarized excitation induces a large Overhauser field ($\bm{B}_{\rm N}$) up to $\sim$0.8 T. Field ${B}_{\rm N}$ is much larger than the measured fluctuation ($\sim$40 mT) and contributes to increase DCP by stabilizing the electron spin. From the excitation power dependences of the Overhauser shift (OHS) and DCP of $X^{+}$-PL, the relation between the Overhauser and Knight fields under zero external magnetic field is discussed. In addition, the depolarization of NSP changes depending on the magnitude of a longitudinal magnetic field (0-1 T), and this allows us to estimate the strength of quadrupolar interaction.

This paper is organized as follows: The InAlAs QD sample and the standard setup for single QD spectroscopy used in this work are described in the next section. In Section III, in order to estimate $\bm{B}_{\rm N}$ from the observed OHS correctly, an evaluation of the electron $g$-factor, including the sign, is carried out. Further, the fluctuation of $\bm{B}_{\rm N}$ is estimated from the correlation between the observed electronic Zeeman splitting and $X^{+}$-DCP. In Section IV, the measurement apparatus is improved to allow simultaneous acquisition of $\sigma^{+}$ and $\sigma^{-}$ components of the PL spectra in order to detect small OHS and DCP  accurately. In Section V, the zero-field DNSP is studied in detail via the excitation power dependences of the OHS and DCP using the improved setup. In addition, the impact of the nuclear quadrupolar effect on DNSP is studied from the dependence of the nuclear depolarization rate on the applied longitudinal magnetic field. Finally, the conclusion of this work is given in Section VI.

\section{Sample and Experimental setup}\label{2nd section}
Self-assembled In$_{0.75}$Al$_{0.25}$As/Al$_{0.3}$Ga$_{0.7}$As QDs grown on an undoped (100)-GaAs substrate by molecular beam epitaxy were used in this work~\cite{Yokoi05,Sasakura04A,Sasakura04B}. The average diameter, height, and density of the QDs were found to be $\sim$20 nm, $\sim$4 nm, and $\sim$5$\times 10^{10}$ cm$^{-2}$, respectively, by the atomic force microscopy measurements of a reference uncapped QD layer. Assuming a lens-shaped QD based on the cross-section transmission electron microscope observation, the number of nuclei in a single QD  is estimated to be roughly $N$$\sim$$ 3 \times 10^{4}$. After the fabrication of small mesa structures (top lateral diameter $\sim$150 nm), the micro-PL measurements in the time-integrated mode were performed at 6 K under longitudinal magnetic fields ($B_{z}$) up to 5 T. 

A continuous wave Ti:sapphire laser was tuned to $\sim$730 nm to provide the transition energy to the foot of the wetting layer of the QDs. The laser beam was focused on the sample surface using a microscope objective lens (M Plan Apo NIR $\times$20, NA$\sim$0.4). The QD emissions were collected by the same objective lens and were detected by a triple-grating spectrometer (Horiba Jobin-Yvon T64000, 1200 grooves/mm$\times$3) and a liquid N$_{2}$-cooled Si-CCD detector (Princeton Instruments Spec-10:100BR). The spectral resolution that determines the PL energies was $\sim$5 $\mu$eV using the spectral fitting. 

Figure~\ref{Fig1}(a) shows typical PL spectra obtained from a target single InAlAs QD at 6 K and 0 T. The PL signals were resolved to linearly polarized components ($\pi^{x}$, $\pi^{y}$). The spectra indicate the exciton family's emissions of this QD: the neutral biexciton ($XX^{0}$), neutral exciton ($X^{0}$), and positive trion ($X^{+}$) from the low energy side. Each charge state could be assigned by considering the fine structure splitting (FSS) and the binding energy. In this figure, the FSS of $\sim$73 $\mu$eV, the inverse pattern of FSS in the $X^{0}$ and $XX^{0}$ peaks, and no splitting in the $X^+$ peak are observed clearly. 

Hereafter throughout this work, we focus on $X^{+}$-PL, which shows the largest peak in this QD. The ground state of $X^{+}$ consists of a spin-singlet formed of two holes and one electron, and thus, the emission polarization is essentially determined only by the electron spin. In this paper, the DCP is given by $\rho_{\rm c}= (I^{-} -I^{+}) / (I^{-} +I^{+})$, where $I^{-(+)}$ is the PL intensity of $\sigma^{- (+)}$ component. Consequently, a high (low) value of the DCP indicates a large (small) degree of electron spin polarization $\langle S_{\rm z} \rangle$ according to  $\langle S_{\rm z} \rangle = \rho_{\rm c}/2$, and the change in $X^{+}$-DCP can be used as a direct measure of $\langle S_{\rm z} \rangle$.

\section{Evaluations of the electron g-factor and fluctuation of the Overhauser field} \label{3rd section}
In this section, the key parameters describing the coupled electron-nuclear spin system, electron $g$-factor ($g^{\rm e}_{\rm z}$), and fluctuation of the Overhauser field ($\Delta B_{\rm N}$), are evaluated from the magneto-PLs under longitudinal magnetic fields $\bm B_{\rm z}$.

First, an individual evaluation of the electron and hole $g$-factors in the $z$-direction ($g_{z}^{\rm e}$, $g_{z}^{\rm h}$) is performed. This procedure is necessary for evaluating $\bm{B}_{\rm N}$ from the observed OHS, which is the energy shift of electron spin state induced by $\bm{B}_{\rm N}$. The method described below is based on the cancellation of $\bm{B}_{z}$ using $\bm{B}_{\rm N}$ on the electron spins.~\cite{Kaji07}

\begin{figure}[tb]
  \begin{center}
    \includegraphics[width=240pt]{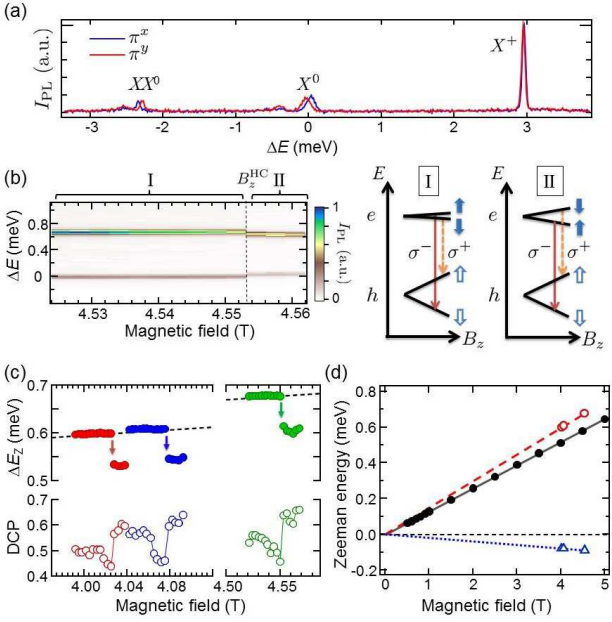}  
  \end{center}
  \caption{(color online) (a) Polarization-resolved  PL spectra of a target InAlAs QD at 6 K and 0 T. The horizontal axis is replotted from the midpoint of the $X^{0}$ doublet. (b) (left panel) Density plot of $X^{+}$-PL as a function of increasing $B_{z}$. The energy axis is replotted from the lower PL peak. (right panel) Level diagrams of the hole (open arrows) and the electron spins (solid arrows) in the smaller (larger) $B_{z}$ than $B_{z}^{\rm{HC}}$, labeled as region I (II). (c) The observed $\Delta E_{\rm Z}$ near three different $B_{z}^{\rm HC}$. The dashed line is the calculated hole Zeeman splitting. In the lower panel, $X^{+}$-DCP is plotted. (d) Observed Zeeman energies of exciton (solid circles), electron (triangles), and hole states (open circles) and the fitting lines.} 
  \label{Fig1}
\end{figure}

The Hamiltonian of HFI between an electron spin $\hat{\bm{S}}$ and $N$-lattice nuclear spins $\hat{\bm{I}}$ can be written as 
\begin{equation}
{\cal H}_{\rm{HF}}=\frac{\nu_{0}}{2}\sum_{j}^{N} A^{j} |\varphi(\bm{r}_{j})|^2 \left[ 2 \hat{I}^{j}_{z} \hat{S}_{z} + \left( \hat{I}^{j}_{+} \hat{S}_{-} + \hat{I}^{j}_{-} \hat{S}_{+} \right) \right] ,
\label{eqHFI}
\end{equation}
where $\nu_{0}$ is the two-atom cell volume, $\bm{r}_{j}$ is the position of the $j$-th nucleus with spin $\hat{I}^{j}$, $\varphi (\bm{r}_j)$ is the normalized electron envelope function, and $A^{j}$ is the coupling constant of HFI~\cite{Abragam}. 
Because a single electron interacts simultaneously with a large number of nuclei, the electron spin experiences an effective magnetic field originated from a mean NSP $\langle \hat{\bm{I}}^{j} \rangle$ expressed as, 
\begin{equation}
\bm B_{\rm N}=\frac{\nu_{0}}{g^{\rm e} \mu_{\rm B}} \sum_{j}^{N} A^{j} |\varphi(\bm{r}_{j})|^{2} \langle \hat{\bm{I}}^{j} \rangle,
\label{eqBN}
\end{equation}
where $\mu_{\rm B}$ is the Bohr magneton ($\sim$58 $\mu$eV/T). Since the HFI on hole spin is usually negligible because of the small coupling constant~\cite{Testelin09,Eble09,Kaji13}, the Zeeman splitting of $X^{+}$-PL under $\bm B_{\rm z}$ is given as $\Delta E_{\rm Z}=g_{z}^{\rm h} \mu_{\rm B} |\bm{B}_{\rm z} |+ g_{\rm z}^{\rm e} \mu_{\rm B} | \bm{B}_{\rm z} + \bm{B}_{\rm N} |$.

The left panel of Fig.~\ref{Fig1}(b) is a density plot of the Zeeman split $X^{+}$-PL under $\sigma^{-}$ excitation with increasing ${B}_{\rm z}$. As clearly shown, the $\Delta E_{\rm Z}$ of $X^{+}$-PL decreases abruptly at the critical magnetic field $B_{\rm z}^{\rm{HC}}$(=4.5515 T). The direction of $\bm{B}_{\rm N}$ generated with $\sigma^{-}$ excitation is opposite to that of $\bm{B}_{\rm z}$ ($\bm{B}_{\rm N} \cdot \bm{B}_{\rm z} <0$). In this situation, the energy mismatch in the e-n FF process is relaxed, and the bistable behavior accompanied by the abrupt change in $B_{\rm N}$ appears. 
The right panel of Fig.~\ref{Fig1}(b) depicts the Zeeman levels of the electron and hole states in smaller (larger) $B_{\rm z}$ than $B_{\rm z}^{\rm{HC}}$, labeled as region I (II). While the electronic Zeeman splitting $\Delta E_{\rm Z}^{\rm e}$ is almost zero in region I, it is revived in region II. This change in $\Delta E_{\rm z}^{\rm e}$ induces the abrupt decrease in $\Delta E_{\rm Z}$ if $g_{\rm z}^{\rm e}$ has the opposite sign to $g_{\rm z}^{\rm h}$. 
Note that $\Delta E_{\rm Z}^{\rm e}$ at $B_{\rm z}= B_{\rm z}^{\rm{HC}}$ is exactly zero because of the full compensation of $\bm{B}_{\rm z}$ by $\bm{B}_{\rm N}$. Thus, the observed $\Delta E_{\rm Z}$ at $B_{\rm z}^{\rm{HC}}$ corresponds to the hole Zeeman splitting, and the hole $g$-factor can be derived as $g_{\rm z}^{\rm h}=\Delta E_{\rm Z}(B_{\rm z}^{\rm{HC}})/\mu_{\rm B}B_{\rm z}^{\rm{HC}}$.

Figure~\ref{Fig1}(c) shows $\Delta E_{\rm Z}$ in the vicinity of three different values of $B_{\rm z}^{\rm HC}$, which can be obtained by changing the excitation power. The increase in the data points of $\Delta E_{\rm Z} (B_{\rm z}^{\rm{HC}})$ helps us to improve the accuracy for determining ($g_{\rm z}^{\rm h}$, $g_{\rm z}^{\rm e}$). Comparing $g_{\rm z}^{\rm h}$ to the exciton $g$-factor $g_{\rm z}^{\rm X}$ ($= g_{\rm z}^{\rm h} + g_{\rm z}^{\rm e}$) that can be obtained with linearly polarized excitation (i.e., $B_{\rm N} =0$), the electron $g$-factor can also be accessed. Note that the $X^{+}$-DCP changes depicted in the lower panel are synchronized with the changes in $\Delta E_{\rm Z}$ (that is, the changes in $B_{\rm N}$). This observation can be described by the effect of $\Delta B_{\rm N}$, as discussed later. 

Figure~\ref{Fig1}(d) shows the Zeeman splittings of the exciton, hole, and electron spins as a function of $B_{\rm z}$. From the linear fittings, the $g$-factors of exciton, hole, and electron spins in this QD were evaluated as $g_{\rm z}^{\rm X}=2.23 \pm 0.01$, $g_{\rm z}^{\rm h} =2.57 \pm 0.01$, and $g_{\rm z}^{\rm e} =-0.34 \pm 0.02$, respectively. These values are close to those reported in previous work~\cite{Kaji07}.

\begin{figure}[tb]
  \begin{center}
    \includegraphics[width=240pt]{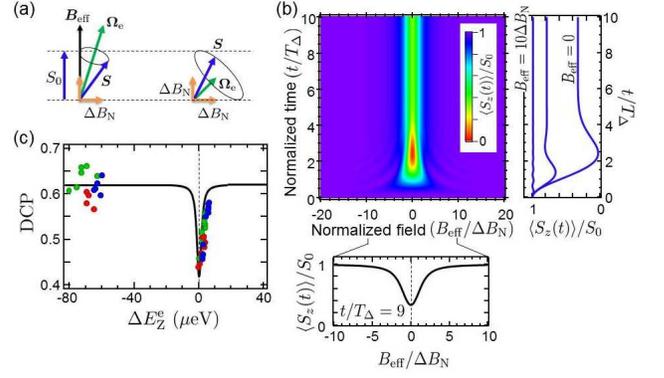}  
  \end{center}
  \caption{(color online) (a) Schematics of electron spin precession around torque vector $\bm{\Omega}_{\rm e}$, which includes $\bm{B}_{\rm{eff}}$ and $\Delta \bm{B}_{\rm N}$. (b) Calculation of $\langle S_{\rm z} (t) \rangle /S_{0}$ as functions of  normalized time ($t/T_{\Delta}$) and magnetic field ($B_{\rm eff}/\Delta B_{\rm N}$). The vertical profiles at $B_{\rm{eff}}/\Delta B_{\rm N} $= 0, 2, 10 are indicated in the right panel, and the horizontal profile at $t/T_{\Delta} $=9 is in the lower panel. (c) Experimentally obtained $X^{+}$-DCP as a function of $\Delta E_{\rm Z}^{\rm e}$.  The solid curve is the calculation obtained using Eq.~(\ref{eqDCP}) with $\tau_{\rm r}$=0.75 ns and $S_{0}$=0.62, which includes the calculation of Fig.~\ref{Fig2}(b) with $T_{\Delta}$=0.8 ns.}
  \label{Fig2}
\end{figure}

\smallskip

Next, we estimate the random fluctuation of the Overhauser field ($\Delta B_{\rm N}$) and the resultant electron spin dephasing time ($T_{\Delta}$). According to the standard picture of spin relaxation, electron spin $\bm{S}$ precesses around the effective magnetic field and loses its coherence via scattering processes. Here, torque vector $\bm{\Omega}_{\rm e}$ is composed of macroscopic field $\bm{B}_{\rm{eff}}$ ($= \bm{B}_{\rm z} +\bm{B}_{\rm N}$) and fluctuating field $\Delta \bm{B}_{\rm N}$. In the absence of $\bm{B}_{\rm{eff}}$, $\bm{S}$ in a QD precesses coherently around $\Delta \bm{B}_{\rm N}$ during its lifetime, which is limited by recombination time $\tau_{\rm r}$. However, electron spin polarization $\langle S_{\rm z} \rangle$, which is an ensemble average over a large number of measurements, decreases within a characteristic time $T_{\Delta}$ due to the random distributions of the direction and magnitude of $\Delta \bm{B}_{\rm N}$, and it converges to $S_{0}/3$ ($S_{0}$: the initial value of $\langle S_{\rm z} \rangle$) over the long time limit (\textit{frozen fluctuation model})~\cite{Merkulov02}. If a large $\bm{B}_{\rm{eff}}$ ($\gg \Delta \bm{B}_{\rm N}$) appears and $\bm{\Omega}_{\rm e}$ is nearly along the $z$-axis, the reduction of $\langle S_{\rm z} (t) \rangle$ is strongly suppressed, as shown in Fig.~\ref{Fig2}(a).

Figure~\ref{Fig2}(b) is $\langle S_{\rm z} (t) \rangle/S_{0}$, calculated as functions of normalized magnetic field $B_{\rm{eff}}/\Delta B_{\rm N}$  and normalized time $t/T_{\Delta}$. The right panel shows the vertical profiles of the figure, which indicates clearly that the increase in $B_{\rm{eff}}$ suppresses the oscillation and reduction of $\langle S_{\rm z} (t) \rangle / S_{0}$. On the other hand, the lower panel is a horizontal plot at $t/T_{\Delta} =9$. In this long time region, the normalized electron spin polarization shows a dip structure centered at the point of zero-$\bm{B}_{\rm{eff}}$, and its width is determined by $\Delta B_{\rm N}$. 

The experimentally obtained DCP is determined by $\langle S_{\rm z} (t) \rangle$ and $\tau_{\rm r}$ ($\sim$0.75 ns), which is evaluated from other independent measurements. Assuming that the orientation of $\Delta \bm{B}_{\rm N}$ is randomly distributed over the accumulation time of the CCD detector (1 s), the DCP of time-integrated $X^{+}$-PL is given by
\begin{equation}
\rho_{\rm c} =\frac{2}{\tau_{\rm r}}\int \langle S_z (t) \rangle \exp(-t/\tau_{\rm r}) dt.
\label{eqDCP}
\end{equation}
Note that this equation leads to a dip structure similar to the one in the lower panel of Fig.~\ref{Fig2}(b), and its width is determined by the ratio of $T_{\Delta}$ to $\tau_{\rm r}$.

Figure~\ref{Fig2}(c) shows the $X^{+}$-DCP as a function of $\Delta E_{\rm Z}^{\rm e}$ obtained from the data set in Fig.~\ref{Fig1}(c). The absence of data points around $\Delta E_{\rm Z}^{\rm e}$=-50-0 $\mu$eV is attributed to the abrupt changes in $\bm{B}_{\rm N}$ and DCP. 
The most important feature is that the observed $X^{+}$-DCP has a dip structure centered at point $\Delta E_{\rm Z}^{\rm e} \sim$0, and it agrees well with the above explanation. By comparing it with the solid curve, which is obtained by Eq.~(\ref{eqDCP}) including the calculated results of Fig.~\ref{Fig2}(b), we can deduce the values $S_{0}=0.62$ and $T_{\Delta}=0.8$ ns.

The magnitude of $\Delta \bm{B}_{\rm N}$ is estimated as $\Delta B_{\rm N} $=40 mT from the relation $\Delta B_{\rm N} = \hbar / (g^{\rm e} \mu_{\rm B} T_{\Delta})$, assuming an isotropic electron g-factor. This value is comparable to those in InAs QDs ($\sim$30 mT)~\cite{Krebs07,Krebs08}, InGaAs QDs ($\sim$10.5 mT)~\cite{Bechtold15}, and InP QDs ($\sim $15 mT)~\cite{Pal07}, and it coincides with the one in a different InAlAs QD~\cite{Kaji12}. Further, the validity of the observed $\Delta B_{\rm N}$ can be confirmed using the QD parameters ($g^{\rm e}$, $\tilde{A}$, $\tilde{I}$, $N$), where $\tilde{A}$ ($\tilde{I}$) is the average of $A^{j}$ ($I^{j}$). 
From the relation $\displaystyle \Delta B_{\rm N} \cong \tilde{A} \tilde{I}/ \sqrt{N} g^{\rm e} \mu_{\rm B}$ with the values ($g^{\rm e}$, $\tilde{A }$, $\tilde{I}$, $N$) = ($-0.34$, 52.6 $\mu$eV, 2.75, $3 \times 10^4$) for our In$_{0.75}$Al$_{0.25}$As QD, $\Delta B_{\rm N}$ is roughly estimated as $\sim $42 mT, which agrees quite well with our observation.

\section{Experimental improvement for DCP measurement}\label{4th section}

Under zero magnetic field, the energy splitting of $X^{+}$-PL is usually less than (or comparable with) the spectral width of QD transitions, and thus, the evaluations of OHS and DCP need to acquire the spectra with ($\sigma^{+}$, $\sigma^{-}$) components separately. Because this standard method is accompanied by more than one (at least two) PL acquisition procedure, the resultant OHS and DCP may be affected by some unfavorable effects such as variation in the excitation density and fluctuation of the sample temperature in the measurement intervals.

\begin{figure}[tb]
  \begin{center}
    \includegraphics[width=240pt]{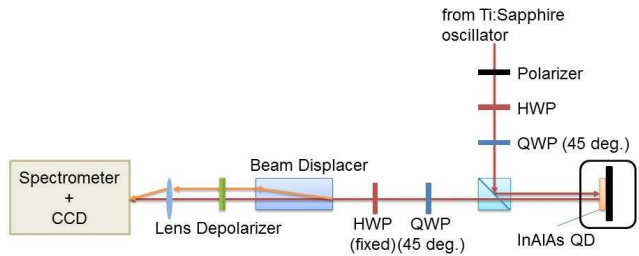}  
  \end{center}
  \caption{(color online) Experimental setup for the fully simultaneous acquisition of $\sigma^+$ and $\sigma^-$ components of the PL spectra for accurate OHS and DCP evaluation.}
  \label{Fig3}
\end{figure}

In order to reduce the measurement errors of OHS and DCP, we improved the experimental setup, as shown in Fig.~\ref{Fig3}; this  enables the simultaneous acquisition of ($\sigma^{+}$, $\sigma^{-}$) components. The ($\sigma^{+}$, $\sigma^{-}$) PL components are converted to ($\pi^{x}$, $\pi^{y}$) ones by a quarter-wave plate (QWP), and they are displaced spatially from each other by a beam displacer, which serves as a linear polarizer in a standard DCP measurement. 
Each displaced PL component is focused on a different detection area of the CCD array. Therefore, the OHS and DCP of the PL can be acquired by a single exposure process. The depolarizer after the beam displacer removes the polarization dependence of the grating diffraction efficiency in the spectrometer. The half-wave plates (HWPs) in the excitation and detection paths are used to compensate for the phase distortion of the circular polarization by many optical elements. Further, in order to avoid the unfavorable effects due to residual magnetization, the He-flow cryostat, which includes the InAlAs QD sample, was replaced apart from the superconducting magnet.

\section{zero-field DNSP and impact of the quadrupolar effect}\label{ZF-DNP}

In a very low magnetic field, the expression of $\bm B_{\rm N}$ is obtained by considering the energy flow and nuclear spin temperature~\cite{OptOrientation,Abragam}, and it can be expressed as
\begin{equation}
\bm B_{\rm N}=b_{\rm n}\frac{\bm B(\bm B \cdot \bm S)}{|\bm B|^2 + \xi  B^2_{\rm L}},
\label{eq4}
\end{equation}
where $b_{\rm n}$ is a proportionality constant, $\xi$ is a coefficient close to unity, and $B_{\rm L}$ ($\sim$0.15 mT for bulk GaAs~\cite{Paget77}) is a small local field due to the dipole-dipole interaction in the nuclear spin ensemble. According to Eq.~(\ref{eq4}), $B_{\rm N}$ cannot be produced without a non-zero magnetic field along $\bm{S}$. 
In general, the DNSP process is performed under an external magnetic field $B_{\rm z}$ that is larger than $B_{\rm L}$, where $B_{z}$ suppresses the nuclear spin relaxation induced by $B_{\rm L}$. However, the strong localization of electron spin and  resultant enhancement of HFI in a QD permit us to produce large $B_{\rm N}$, even under zero-$B_{\rm z}$.  
\begin{figure}[tb]
  \begin{center}
    \includegraphics[width=240pt]{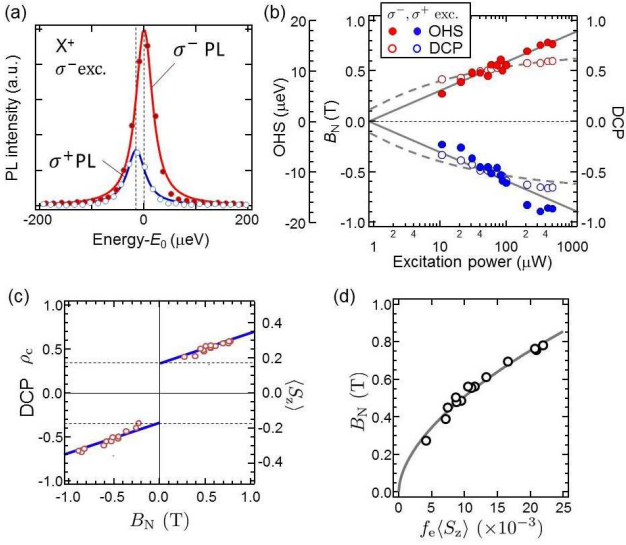}  
  \end{center}
  \caption{(color online) (a) $\sigma^{+}$ and $\sigma^{-}$ components of $X^{+}$-PL spectra under $\sigma^-$-excitation at 0 T. For clarity, the spectra are shifted by energy $E_{0}$=1.6411 eV. An OHS of $\sim$15 $\mu$eV is observed. (b) Excitation power dependences of OHS ($\propto B_{\rm N}$) (solid circles) and DCP (open circles). (c) The measured DCP (=$2 \langle S_z \rangle$) is plotted as a function of $B_{\rm N}$. 
(d) Obtained relation between $B_{\rm N}$ and $f_{\rm e}\langle S_{\rm z} \rangle$, which corresponds to the relation between $B_{\rm N}$ and $B_{\rm K}$.}
  \label{Fig4}
\end{figure}

Figure~\ref{Fig4}(a) shows the $X^{+}$-PL under $\sigma^{-}$ excitation at $B_{\rm z} =0$. The spectra were obtained with the setup described in Section~\ref{4th section}. The sharp spectra in the figure indicate the full width at half maximum (FWHM) of $\sim$40 $\mu$eV and an OHS of 15 $\mu$eV. 
This OHS corresponds to $B_{\rm N}$=760 mT using the obtained value $g_{\rm z}^{\rm e}=-0.34$. In addition, the DCP is evaluated as $\sim$60\%, which is larger than the value expected from the \textit{frozen fluctuation model}, as discussed in Section III. The reason for the high DCP here is the presence of a large $B_{\rm N}$, which is sufficient to suppress the electron spin relaxation by $\Delta B_{\rm N}$ ($\sim$40 mT) and stabilize $\langle S_{z} \rangle$. 
This large $B_{\rm N}$ under zero external magnetic field can be formed via nuclear spin cooling by the Knight field~\cite{OptOrientation,Lai06}. 

Because HFI is reciprocal between the electron and nuclear spin systems, the nuclei are also affected by an effective magnetic field $\bm B_{\rm K}$ (Knight field) induced by the average electron-spin polarization $\langle \bm {\hat S} \rangle$. The time-averaged Knight field acting on one specific nucleus $j$ is given by
\begin{equation}
\bm B_{\rm K}^j=f_{\rm e} \frac{\nu_0 A^j}{g_{\rm N}\mu_{\rm N}}|\varphi(\bm r_j)|^2 \langle \bm {\hat S} \rangle
=f_{\rm e} b_{\rm e}^j \langle \bm {\hat S} \rangle,
\label{eq5}
\end{equation}
where $f_{\rm e}$ is a filling factor representing the occupation of a QD by an unpaired electron $(0 \le f_{\rm e} \le 1$), $g_{\rm N}$ is the nucleus g-factor, $\mu_{\rm N}$ is the nuclear magneton, and $b_{\rm e}^j$ is a proportionality constant. The Knight field can be tuned experimentally  by changing excitation power $P_{\rm{ex}}$ and the polarization of the excitation light; the former affects $\bm{B}_{\rm K}$ through the change in $f_{\rm e}$, and the latter does so through the change in $\langle \bm{\hat{S}} \rangle$.

In order to investigate the role of $\bm{B}_{\rm K}$ in the process of nuclear spin cooling, we measured the $P_{\rm ex}$-dependences of OHS and DCP, as shown in Fig.~\ref{Fig4}(b). A slight asymmetry between $\sigma^-$ and $\sigma^+$ excitations may be caused by the variation of the excitation power. The OHS increases almost linearly on $\log$($P_{\rm ex}$) in the experimental region of $P_{\rm ex}$, while the corresponding DCP saturates at 60\% in the high-$P_{\rm{ex}}$ region. The change in DCP is due to the stabilization of $\langle S_{\rm z} \rangle$ by $\bm B_{\rm N}$. 
To highlight this scenario, we plotted DCP versus $B_{\rm N}$ (Fig.~\ref{Fig4}(c)) using the data set of Fig.~\ref{Fig4}(b), where $|\rho_{\rm c}|$ increases almost linearly with $| \bm{B}_{\rm N}|$. Even at the minimum $P_{\rm{ex}}$ in this experiment, a $B_{\rm N}$ of $\sim$200 mT, which is larger than $\Delta B_{\rm N}$, is produced, and thus, an obvious reduction of DCP ($\rho_{\rm c}=2 \langle S_{\rm z} \rangle$) caused by $\Delta B_{\rm N}$ does not appear in the figure. 

In the case of QDs where large nuclear fields are generated, the dependence of the e-n FF rate (hidden in $b_{\rm n}$ in Eq.~(\ref{eq4})) on the total magnetic field $\bm{B} + \bm{B}_{\rm N}$ has to be taken into account. Thus, the expression for $\bm{B}_{\rm N}$ (Eq.~(\ref{eq4})), is replaced by the steady state solutions of the following rate equation for $\langle I_{\rm z} \rangle$:
\begin{equation}
\frac{d \left< I_{\rm z} \right>}{dt} = -\frac{1}{T_{\rm {NF}}} \left[\left< I_{\rm z} \right> - \tilde{Q} \left( \left< S_{\rm z} \right> -\left< S_{\rm z} \right>_0 \right)  \right] -\frac{1}{T_{\rm{ND}}} \left< I_{\rm z} \right>,
\label{eq6}
\end{equation}
where $\tilde{Q}=\sum_j I^j(I^j+1)/[N S(S+1)]$ is a numerical constant representing the angular momentum conversion and $\left< S_{\rm z} \right>_{0}$ is the average electron spin polarization at thermal equilibrium. The first term on the right-hand side represents the DNSP formation driven by the non-equilibrium electron spin, and the second term accounts for the depolarization of nuclear spin system with time constant $T_{\rm{ND}}$. 
The DNSP formation rate $1/T_{\rm NF}$ is given as
\begin{equation}
\frac{1}{T_{\rm NF}}=2 f_{\rm e} \tau_{\rm c} \left(\frac{\tilde{A}}{N \hbar} \right)^2 \bigg/ \left[{1+\left(\frac{\tau_{\rm c}}{\hbar}\right)^2({\rm g}_{\rm z}^{\rm e} \mu_{\rm B} B_{\rm z} \pm \tilde{A} \langle I_{\rm z} \rangle)^2}\right],
\label{eq7}
\end{equation}
where $\tau_{\rm c}$ is the correlation time of the hyperfine perturbation. Using Eqs.~(\ref{eq6}) and (\ref{eq7}), $f_{\rm e}$ can be estimated from the fitting of the data in Fig.~\ref{Fig4}(b), and $f_{\rm e}$ was found to be proportional to $\log$($P_{\rm ex}$) (not shown here).

In Fig.~\ref{Fig4}(d), $| \bm{B}_{\rm N} |$ is plotted as a function of the product of $f_{\rm e}$ and $\langle S_{\rm z} \rangle$. From Eq.~(\ref{eq5}), $f_{\rm e} \langle S_{\rm z} \rangle$ is considered to be proportional to the magnitude of $\bm{B}_{\rm K}$. 
Because the constant $b_{\rm e}$, which is the average of $b_{\rm e}^{j}$, of our QD is thought to be $\sim$40 mT, the magnitude of $B_{\rm K}$ is estimated to be $0.2-1.0$ mT in the experimental region of $P_{\rm ex}$; this effective field for the nucleus is larger than $B_{\rm L}$. 
Actually, Lai \textit{et al}. evaluated $B_{\rm K}=\pm0.6$ mT directly through the field compensation (i.e., $\bm B_{\rm K} + \bm B_{\rm z} =0$) accompanied by the reduction of X$^{-}$-DCP in InAs QDs under $\sigma^{\pm}$ excitation~\cite{Lai06}.  
From the relation $g_{\rm e}\mu_{\rm B} \bm B_{\rm N}^{\rm max}=N g_{\rm N}\mu_{\rm N} \bm B_{\rm K}^{\rm max}$,  $|\bm B_{\rm K}^{\rm max}|$ is estimated to be a few tens of millitesla.

\begin{figure}[tb]
  \begin{center}
    \includegraphics[width=240pt]{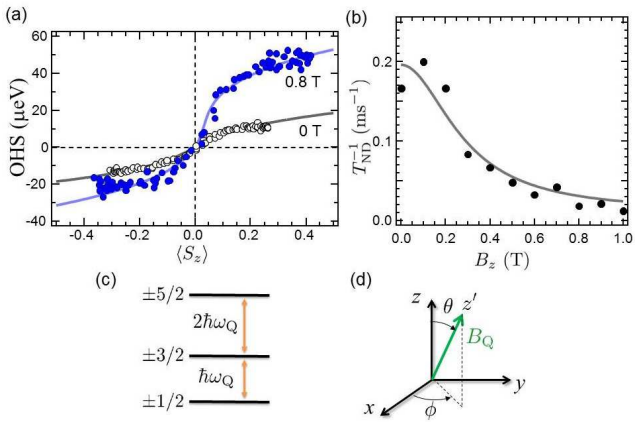}  
  \end{center}
  \caption{(color online) (a) OHS as a function of $\langle S_{\rm z} \rangle$ at $B_{\rm z}=$0 T and 0.8 T ($P_{\rm exc}$=100 $\mu$W). $\langle S_{\rm z} \rangle$ is converted from the measured DCP from $X^{+}$ PL line intensity. Solid lines are the fitting curves using Eq.~(\ref{eq6}). (b) Dependence of nuclear spin relaxation rate $T_{\rm ND}^{-1}$ on longitudinal magnetic field $B_{\rm z}$. (c) Energy level diagram at $B_{\rm z}$=0 under the influence of QI for nuclear spin $I$=5/2. (d) Quadrupolar field $B_{\rm Q}$ is tilted by $\theta$ to the $z$ axis. }
  \label{Fig5}
\end{figure}
Finally, we study the role of the quadrupolar field and evaluate the magnitude experimentally. 
Figure~\ref{Fig5}(a) indicates the OHS as a function of the measured $\langle S_{\rm z} \rangle$ under the external magnetic fields $B_{\rm z}=0$ and 0.8 T.
The excitation power was fixed to 100 $\mu$W, and the polarization of the excitation light was changed systematically. While the change in OHS at 0 T is exactly antisymmetric with respect to the change in the sign of $\langle S_{\rm z} \rangle$, this is not the case at 0.8 T; the OHS in positive $\langle S_{\rm z} \rangle$ is larger than that in negative $\langle S_{\rm z} \rangle$, which is caused by the difference of the DNSP formation rate.
Note that the application of $B_{\rm z}$ leads to an increase in OHS even in the negative $\langle S_{\rm z} \rangle$ region. This observation seems to be strange at first view because the generated $\bm{B}_{\rm N}$ has the same direction with $\bm{B}_{\rm z}$ in the negative $\langle S_{\rm z} \rangle$ region. According to Eq.~(\ref{eq7}), $1/T_{\rm{NF}}$ and the resultant OHS decrease if other parameters are unchanged.

Solid curves in the figure are the fittings obtained from Eq.~(\ref{eq6}) by changing $\langle S_{\rm z} \rangle$, where the nuclear spin depolarization rate $1/ T_{\rm{ND}}$ is unique for each $B_{\rm z}$. The best fittings show that $T_{\rm ND}=6$ and 56 ms for $B_{\rm z}$=0 and 0.8 T, respectively. The result indicates that the application of $B_{\rm z}$ suppresses $1/T_{\rm ND}$. We then investigated the $B_{\rm z}$-dependence of $1/T_{\rm{ND}}$ in detail by changing the magnitude of $B_{\rm z}$ up to 1 T while other experimental conditions were unchanged.
Figure~\ref{Fig5}(b) depicts the deduced $1/T_{\rm{ND}}$ as a function of $B_{\rm z}$. As clearly shown, the nuclear depolarization rate decreases systematically with increasing $B_{\rm z}$ and approaches a saturated value.

The following two mechanisms may contribute to the nuclear depolarization term: the dipolar interaction between nuclear spins and the quadrupolar interaction (QI). It is widely accepted that the dipolar interaction is responsible not only for a fast depolarization in a very low magnetic field ($\le B_{\rm L}$) but also for a slow depolarization through the nuclear spin diffusion process~\cite{Adachi12}. Although the nuclear spin diffusion process is effective even under a large $B_{\rm z}$, the efficiency of this depolarization mechanism does not depend on the magnitude of $B_{\rm z}$. Therefore, the observed change in $1/T_{\rm{ND}}$ seems to originate from the QI mechanism.

The QI originates from the coupling between an electric quadrupolar moment of nuclear spin ($I >1/2$) and an inhomogeneous electric field gradient (EFG) caused by the alloy disorder and/or the local strain in the crystal, and it splits the $(2I+1)$ spin levels according to the square of their angular momentum projection, as depicted in Fig.~\ref{Fig5}(c). Here, the splitting energy is characterized by the term $\hbar \omega_{\rm Q}$, which is proportional to the nuclear quadrupolar moment and the EFG. 

For the purpose of quantitative comparison, we introduce an effective magnetic field by the QI, $B_{\rm Q} \equiv \hbar \omega_{\rm Q} / g_{\rm N} \mu_{\rm N}$. Further, the EFG in a QD is assumed to have cylindrical symmetry with respect to the $z$-axis, as shown in Fig.~\ref{Fig5}(d) for simplicity. The influences of QI in the self-assembled QDs are thought to be large compared to those in unstrained systems because of a large residual strain associated with the QD formation process, and the estimated value of $B_{\rm Q}$ is in about the 100 mT range. This effective field, originated from QI, suppresses the dipolar coupling and resultant nuclear depolarization process effectively.

The EFG should orient almost along the $z-$axis, which has been confirmed experimentally in In(Ga)As QDs by single-QD NMR~\cite{Chekhovich12,Chekhovich15}. However, if the EFG axis is tilted by an angle $\theta$, the QI may be responsible for a specific mechanism of nuclear spin depolarization. Under this condition, one nuclear state $| I_{\rm z} \rangle$ couples with the other states $| I_{\rm z} \pm 1 \rangle$, and thus the differences in the populations of these  states and resultant polarizations are partially canceled out. Further, a QI with non-zero $\theta$ may also contribute to $1/ T_{\rm{ND}}$ by coupling with the temporal fluctuation of the longitudinal part of HFI (i.e., $\propto S_{\rm z} I_{\rm z}$) as suggested by Huang \textit{et al}~\cite{Huang10}. This type of depolarization seems to be suppressed by applying a longitudinal magnetic field because $B_{\rm z}$, which is comparable or larger than $B_{\rm Q}$, can restore the eigenaxis of nuclear spin along $z$ and decouple between $| I_{\rm z} \rangle$ and $| I_{\rm z} \pm 1 \rangle$.

From these considerations, the nuclear depolarization rate can be assumed as follows~\cite{Urbaszek13,Krebs08}:
\begin{equation}
T_{\rm ND}^{-1}=T_{\rm ND \infty}^{-1}+T_{\rm ND0}^{-1}\left[ 1+\left(\frac{B_{\rm z}}{B'_{\rm Q}} \right)^2 \right]^{-1},
\label{eq8}
\end{equation}
where $1/T_{\rm ND \infty}$ is a constant term representing the depolarization rate at the high-$B_{\rm z}$ limit, $1/T_{\rm ND0}$ is an amplitude of the Lorentzian part describing qualitatively the slowdown of $1/T_{\rm{ND}}$ at high-$B_{\rm z}$, and $B'_{\rm Q}$ is a measure of the quadrupolar field. 
The solid curve in Fig.~\ref{Fig5}(b) is a fitting obtained using Eq.~(\ref{eq8}), and it reproduces the experimental result quite well.
From the fitting, $T_{\rm ND \infty}=100$ ms, $T_{\rm ND0}=5$ ms, and $B'_{\rm Q}=280$ mT were obtained. The obtained value of $B'_{\rm Q}$ is comparable with those evaluated in self-assembled In(Ga)As/GaAs QDs ($\sim$400 mT~\cite{Krebs08} and $\sim$300 mT~\cite{Maletinsky09}).

When a QD does not include any charge carriers, the dipolar interaction between nuclear spins plays a dominant role in the decay of NSP. In particular, the nuclear spin diffusion is still effective even under a large $B_{\rm z}$, and thus, it induces a severe problem related to the \textit{nuclear spin bath noise}. Along this line, the quadrupolar interaction has a positive effect thanks to the resultant non-equivalent energy spacing of nuclear spin levels; it inhibits the simultaneous spin flip between nuclear spins due to the energy mismatch. In contrast, the quadrupolar interaction can induce a specific mechanism of nuclear spin depolarization if the QD is occupied with an electron spin and the principal axis of EFG is away from the $z$-axis, as discussed above. These properties make QI more important and interesting for understanding the coupled electron-nuclear spin system deeply, and thus, our observation will serve as useful information.

\section{Conclusion} 
We investigated the DNSP in a self-assembled InAlAs quantum dot under a zero external magnetic field by focusing on the PL from the positive trion $X^+$. The DCP of $X^{+}$-PL works as a direct measure of the average electron spin polarization $\langle S_{\rm z} \rangle$. 
 First, the key parameters describing the coupled electron-nuclear spin system, electron g-factor and fluctuation of the Overhauser field of a target single QD were evaluated experimentally. After the experimental setup was improved to enable highly accurate evaluations of OHS and DCP, the zero-field DNSP was studied in detail. 
The spin-selectively-excited electron generated considerable Overhauser fields of up to $\sim$0.8 T. 
From the excitation power dependences of OHS and DCP, the relation between the generated Overhauser field and generating Knight field at 0 T was clearly obtained. The resulting comprehensive knowledge about zero-field DNSP, including the evaluations of key parameters for a typical self-assembled QD, will serve as useful information.

Further, we found a gradual reduction by almost one-order of the nuclear depolarization rate $T_{\rm{ND}}^{-1}$  by increasing the longitudinal magnetic field up to 1 T. The change in $T_{\rm{ND}}^{-1}$ seemed to be related to the quadrupolar interaction with the tilted principal axis, and this observation can be interpreted as the restoring of the eigenaxis of nuclear spin. We believe that this observation in zero-field DNSP has not been previously reported and can contribute to the engineering of nuclear spins in QDs.

This work was supported in part by JSPS KAKENHI (Grants No. 25247047 and No. 26800162).

\end{document}